\documentclass[aps,prl,twocolumn,groupedaddress,a4paper,10pt,floatfix]{revtex4-1}

\usepackage{epsfig}
\usepackage{latexsym}
\usepackage{amsmath}
\usepackage{amssymb}
\usepackage{amsfonts}
\usepackage{graphicx}
\usepackage[left]{lineno}
\usepackage{url}
\usepackage{nicefrac}
\usepackage[normalem]{ulem}

\begin{document}

\title{A Trap Model for Clogging and Unclogging in Granular Hopper Flows}

\author{Alexandre Nicolas}
\email{alexandre.nicolas@polytechnique.edu}
\affiliation{LPTMS, CNRS, Univ. Paris-Sud, Universit´e Paris-Saclay, 91405 Orsay,
France.}

\author{\'{A}ngel Garcimart\'{\i}n}
\affiliation{Departamento de F\'{\i}sica y Matem\'{a}tica Aplicada, Facultad de Ciencias,
Universidad de Navarra, Pamplona, Spain}

\author{Iker Zuriguel}
\affiliation{Departamento de F\'{\i}sica y Matem\'{a}tica Aplicada, Facultad de Ciencias,
Universidad de Navarra, Pamplona, Spain}

\date{\today}

\begin{abstract}
Granular flows through narrow outlets may be interrupted by the formation of arches or vaults that clog the exit. These clogs
may be destroyed by vibrations. A feature which remains elusive is the broad distribution $p(\tau)$ of clog lifetimes $\tau$ measured under constant vibrations. Here, we propose a simple model for arch-breaking, in which the
vibrations are formally equivalent to thermal fluctuations in a Langevin equation; the rupture of an arch corresponds to the escape from an energy trap.
We infer the distribution of trap depths from experiments and, using this distribution, we show that the model captures the empirically observed heavy tails in $p(\tau)$. 
These heavy tails flatten at large $\tau$, consistently with experimental observations
under weak vibrations, but this flattening is found to be systematic, thus questioning the ability of gentle
vibrations to restore a finite outflow forever. The trap model also replicates recent results on the effect of increasing gravity on
the statistics of clog formation in a static silo. Therefore, the proposed framework points to a common physical underpinning to the processes of clogging and unclogging, despite their different statistics.
\end{abstract}

\maketitle

When discrete bodies flow through a constriction, there exists a risk of clogging, due to the spontaneous formation of arch-like (in two dimensions) or dome-like
(in three dimensions) structures obstructing
the bottleneck. This phenomenon can arise in an impressive variety of systems
 \cite{helbing2000simulating,to2001jamming,delarue2016self,haw2004jamming,genovese2011crystallization} and similar features 
have been observed in most of them, from granular
packings flowing out of a vibrating silo \citep{janda2009unjamming}
and colloids flowing through an orifice under a pressure gradient \cite{hidalgo2017colloids}, to living beings, such as mice \cite{lin2016experimental}, sheep \cite{zuriguel2014clogging}, and pedestrians \cite{pastor2015experimental}. In particular,
while the flow intervals $t_f$ between clogs are exponentially distributed, the distribution of lifetimes $\tau$ of (temporary) individual clogs
%(i.e. the time that it takes from the formation of a blocking structure until its destruction)
is heavy-tailed and can be fitted to a power law, viz., $ p(\tau) \sim \tau^{-\alpha}$. When the exponent $\alpha$ is smaller than 2, 
the average clog lifetime $\left\langle \tau\right\rangle$ does not converge; the mean outflow thus vanishes, which in practice means that
extremely long clogs will dominate the process. This defines the clogged regime \cite{zuriguel2014clogging}. In contrast, a finite mean 
outflow is obtained for $\alpha>2$, despite the flow intermittency.

In granular hopper flows, the unclogged regime $\alpha>2$ can be reached by enlarging the outlet or by applying stronger vibrations to the setup, 
both of which lead to larger values of $\alpha$, hence fewer long-lived clogs \cite{zuriguel2014clogging}. It is still debated
whether clogs completely disappear above a critical outlet size
in the absence of vibrations, or whether an (infinite) static silo will always clog up, \emph{eventually} \citep{zuriguel2005jamming,to2005jamming,thomas2013geometry,thomas2015fraction}.
Beyond this conceptual question, differences have been put in the limelight between the static case and the
shaken one. In particular, the formation of a clog and its destruction through
vibrations follow different statistics, which has suggested that these processes are fundamentally distinct \cite{to2017flow,janda2009unjamming,zuriguel2017clogging,mankoc2009role}. Indeed,  clogging
is described as a Poissonian process characterized by a constant probability of formation of a stable arch.
On the other hand, the unclogging probability is not constant over time: The longer an arch has survived, the longer it 
will probably still live. Accordingly, 
this phenomenon was ascribed to aging \cite{blanc2014and}, but so far this explanation has not been confirmed
by experimental evidence.

In this Letter we promote a different explanation, centered on the heterogeneous \emph{native} arch stabilities. We put forward a simple model that rationalizes the heavy tails of the unclogging process in vibrated silos. Remarkably, when applied to static silos, the model is found to reproduce several characteristic features, thus hinting at a common
%physical
underpinning for clogging and unclogging.

\begin{figure}[h]
\centering
\includegraphics[width=0.95\columnwidth]{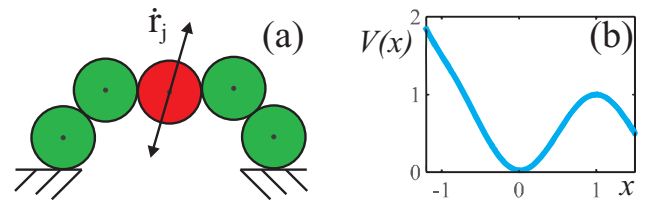}
\caption{
\label{fig:sketch_arch}\textbf{(a)} Sketch of an arch of vibrated grains.  \textbf{(b)} Profile of the potential well used in the numerical simulations.}
\end{figure}

Consider the arch sketched in Fig.~\ref{fig:sketch_arch}a, which is subjected to vertical vibrations characterized by a
 dimensionless acceleration $\Gamma=\frac{ A \omega^2}{g}$, where $A$ is the amplitude and $\omega$ is the angular frequency. Newton's equation of motion for grain \emph{j} (at position $\boldsymbol{r}_{j}$) reads
\begin{equation*}
\boldsymbol{\ddot{r}}_{j}=-\frac{\partial V}{\partial\boldsymbol{r}_{j}}\left(\boldsymbol{r}_{1},\ldots,\boldsymbol{r}_{N}\right)+\boldsymbol{g}+\boldsymbol{f}_{j}+\xi(t) \label{eq:eq1}
\end{equation*}
where the mass of the grain has been set to one. The first term on the
right-hand side accounts for the conservative interactions between
grains ($V$ is the global potential energy), $\boldsymbol{g}$ is
the gravity, and $\boldsymbol{f}_{j}$ is a dissipative
frictional force. The vibrations induce an extra force $\xi(t)$. Let us focus on the weakest link $j$ in the
arch and overlook the deformation of the rest of the arch, whence
we approximate $V\left(\boldsymbol{r}_{1},\ldots,\boldsymbol{r}_{N}\right) \approx V\left(\boldsymbol{r}_{j}\right)$. This is supported by experimental observations indicating that the particle with the largest angle dominates the breaking process \cite{lozano2012breaking}. To simplify the picture further, the gravitational potential energy is included in $V$ and we assume quasi-one dimensional motion ($\boldsymbol{r}\rightarrow x$), viz., $\ddot{x} = -V^{\prime} \left(x\right)+f+\xi(t)$ where the $j$ subscripts have been dropped. As vibrations are symmetric, $\left\langle \xi(t)\right\rangle =0$. During the clogging event, $\left\langle \ddot{x}\right\rangle \approx0$ and $V^{\prime}(x)$ evolves much more slowly than $\xi(t)$, so taking
the variance of the equation of motion over a small time window yields
$\left\langle \ddot{x}^{2}\right\rangle \approx\left\langle \xi^{2}\right\rangle$. Here, we have also assumed that $f$ increases
at most linearly with the \emph{velocity} $|\dot{x}|$ of the grain, so that at high $\omega$, $\left\langle \ddot{x}^{2}\right\rangle \sim \omega^{2}\left\langle \dot{x}^{2}\right\rangle \gg\left\langle f^{2}\right\rangle $. Finally, we note that the acceleration of the grain must be roughly proportional to the acceleration measured on the vibrated setup, viz., $\left\langle \ddot{x}^{2}\right\rangle =\mathrm{Tr}(\omega,\rho,\ldots)\Gamma^{2}$, where the transmission factor $\mathrm{Tr}$ includes the dependence on the material parameters of the grain, the frequency $\omega$, the density $\rho$, and so on. To leading order, overlooking these dependencies and the possible temporal correlations of the vibrations $\xi(t)$, we arrive at

\begin{equation}
\ddot{x}=-V^{\prime}\left(x\right)+f+\xi(t) \label{eq:NewtonSxi}
\end{equation}
with $\left\langle \xi\left(t\right)\right\rangle  = 0 $ and $\left\langle \xi\left(t\right)\xi\left(t^{\prime}\right)\right\rangle \propto\Gamma^{2}\delta\left(t-t^{\prime}\right)$.
We notice that, in the case of viscous friction, i.e., $f\equiv-\gamma\dot{x}$ (with $\gamma$ the drag coefficient), Eq.~\ref{eq:NewtonSxi} is a Langevin 
equation with a vibration-induced Gaussian white noise associated with a temperature $\beta^{-1}=\Gamma^{2} / \gamma$.
In the present work, we focus on this analytically tractable case, leaving for a separate study its generalization to other models of friction. The stability of the arch 
implies that $x$ sits in a basin of $V$, constrained by an energy barrier of height, say, $E_{b}$. The hopping rate over such a barrier was worked out by Kramers \cite{Kramers1940} and reads

\begin{equation}
k\equiv\left\langle \tau\right\rangle ^{-1}=\nu e^{-\beta E_{b}}, \label{eq:Kramers_formula}
\end{equation}
where the attempt frequency $\nu$ depends on $\gamma$ and
%implicitly includes the dependence on $E_b$ in
on the angular vibrational frequencies $\omega_{0}$ and $\omega_{b}$ at the minimum and at the saddle point.
Kramers' formula holds in the moderate to high damping regime $\gamma>\nu$, for $\beta E_{b}\ll1$ (hence, $k<\nu$).
%In the strong damping limit, $\nu\simeq\frac{\omega_{0}\omega_{b}}{\pi\gamma}$.
For a single $E_b$, hence a single $k$, different realizations of the noise yield an exponential distribution of escape times $\tau$ \cite{hanggi1990reaction}

\begin{equation}
p(\tau|k)=k\,e^{-k\tau} \label{eq:dist_t_exp}
\end{equation}

In reality, energy barriers are expected to take a whole range of values, reflected by a distribution $p(E_b)$. So will then the hopping rates $k$, by virtue of 
$p(k)dk = p(E_{b})dE_{b}$. In this situation, the distribution of escape times $\tau$ ($\tau\geqslant \nu^{-1}$)
is given by the convolution 
\begin{equation}
p(\tau)=\int_{0}^{\nu}dk\,p(\tau|k)\,p(k). \label{eq:p_convolution}
\end{equation}

The remaining step is to gather information on the features of the energy landscape, and more specifically $p(E_{b})$. 
To do so, we exploit the arch-destabilization experiments performed by Lozano et al. \cite{lozano2012breaking}, in which an acceleration ramp $\Gamma(t)=\dot{\Gamma}\, t$ was applied to the arch until it breaks.
An arch will typically break at an intensity $\Gamma_{c}$ such that the breaking time $\left\langle \tau\right\rangle $ is of the order of the experimental
ramp time ($\propto \dot{\Gamma}^{-1}$). Assimilating $\tau$ to the escape time from a trap of depth $E_{b}$ and using Eq.~\ref{eq:Kramers_formula} with $\beta=\frac{\gamma}{\Gamma^{2}}$, we get $E_{b} \approx \ln( \nu / \dot{\Gamma} )\, \Gamma_{c}^{2} / \gamma$. 
Neglecting the weak (logarithmic) dependences on $\nu$ and $\dot{\Gamma}$ , we arrive at 
\begin{equation}
E_{b} \approx \frac{\Gamma_{c}^{2}}{\gamma}.  \label{eq:Eb_from_Gammac}
\end{equation}
More rigorous arguments
\footnote{These arguments are exposed in the Supplemental Material} lead to the same scaling. Besides, it was observed that 
the average value $\Gamma_{c}$ at which arches broke was virtually insensitive to $\dot{\Gamma}$ in a given range \cite{lozano2012breaking}. 
Equation~\ref{eq:Eb_from_Gammac} implies that the exponential distributions
$p\left(\Gamma_{c}\right)$ measured experimentally for all tested outlet sizes $D$ \cite{lozano2015stability}
translate into a Weibull distribution of energy barriers $p(E_{b})$, viz.,
\begin{equation}
\label{eq:D(Eb)}
E_{b}=E_{b}^{\star}y^{a}\text{ with }p(y)=e^{-y}
\end{equation}
where $a=2$, $y \equiv \frac{\Gamma_{c}}{\left\langle \Gamma_{c}\right\rangle }$ and $E_b^\star$ implicitly depends on $D$.

\begin{figure*}
\includegraphics[width=1.5\columnwidth]{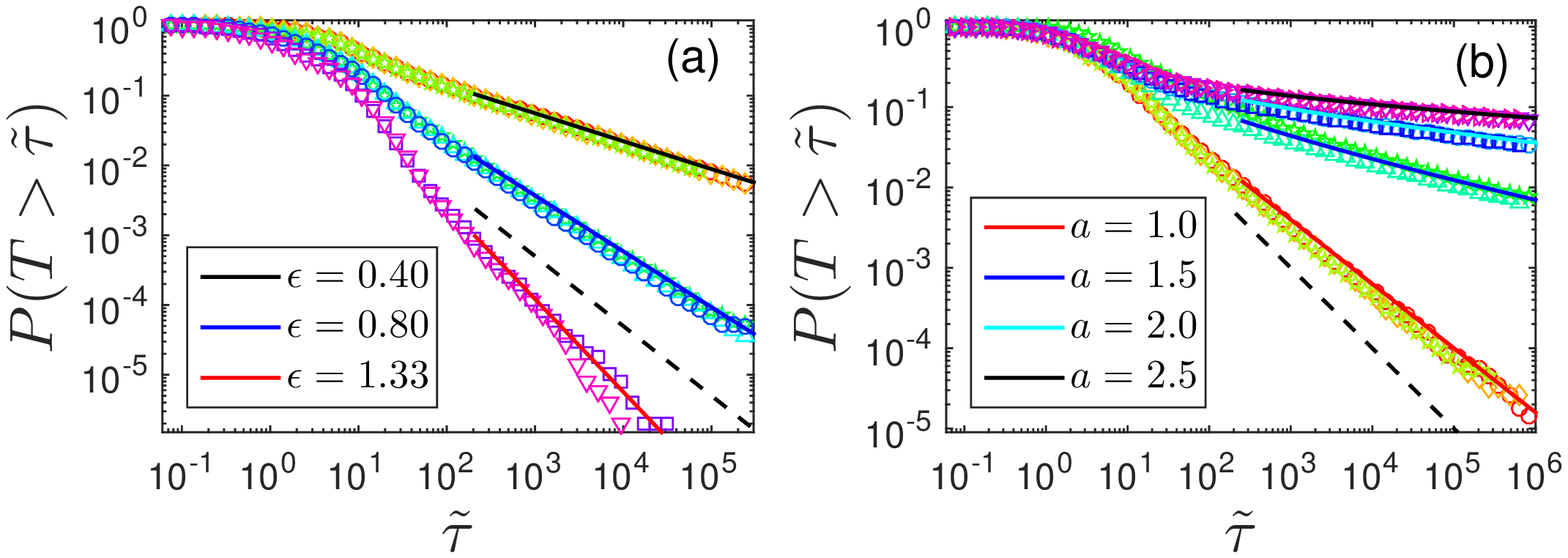}
\caption{\label{fig:CCDF_alphas} \textbf{(a)} Survival functions $P(T>\tilde{\tau})$ for an exponential distribution of barrier heights ($a=1$) as a function of rescaled time $\tilde\tau$, for the values of the vibrational temperature $\epsilon$ indicated in the legend. \textbf{(b)} Survival functions $P(T>\tilde{\tau})$ at fixed $\epsilon=0.8$, for different shape parameters $a$ versus $\tilde\tau$. In both plots, the different symbols refer to various parameter sets $(\gamma,\Gamma,E_{b}^{\star})$. The data collapse onto  master curves coinciding with the predictions of Eqs.~\ref{eq:Pcum_a=1}-\ref{eq:CCDF_general_approximation}, shown as thick lines. The dashed lines in black have slope -1.}
\end{figure*}

For more generality, we will nonetheless study Weibull distributions $p(E_b)$ of arbitrary inverse shape parameters $a$.
We start the discussion with the simple case $a=1$, i.e., an exponential distribution. Equation~\ref{eq:p_convolution}
then turns into
\begin{equation*}
p(\tau) = \epsilon\int_{0}^{\nu}dk\,e^{-k\tau}\left(\frac{k}{\nu}\right)^{\epsilon},
\end{equation*}
where the dimensionless temperature $\epsilon\equiv\frac{\Gamma^{2}}{\gamma E_{b}^{\star}}$ has been introduced.

Changing variables to $\tilde{k}\equiv k\tau$ and
rescaling time as $\tau\rightarrow\tilde{\tau}\equiv\nu\tau$,
one easily arrives at the pdf for $\tilde{\tau}$,
\begin{equation*}
p(\tilde{\tau}) =  \frac{\epsilon}{\tilde{\tau}^{1+\epsilon}}\int_{0}^{\tilde{\tau}}d\tilde{k}\,e^{-\tilde{k}}\tilde{k}^{\epsilon}.
\end{equation*}
The complementary cumulative distribution function  (CCDF) $P(T>\tilde{\tau})$, also called survival function, then reads

\begin{eqnarray}
&& \int_{\tilde{\tau}}^{\infty}p(T)dT  = \epsilon\int_{0}^{\infty}dk\,e^{-k}k^{\epsilon}\int_{\max\left(k,\tilde{\tau}\right)}^{\infty}dT\,T^{-1-\epsilon} \nonumber \\
&& = \left(\int_{0}^{\tilde{\tau}}dk\,e^{-k}k^{\epsilon}\right)\tilde{\tau}^{-\epsilon}+e^{-\tilde{\tau}}  \approx \mathit{\Gamma}\left(1+\epsilon\right)\tilde{\tau}^{-\epsilon}
\label{eq:Pcum_a=1}\end{eqnarray}
where we have introduced the Gamma function $\mathit{\Gamma}$, so
\begin{equation*}
p(\tilde{\tau})\approx\epsilon\mathit{\Gamma}\left(1+\epsilon\right)\tilde{\tau}^{-1-\epsilon}\text{ for }\tilde{\tau}\rightarrow\infty.
\end{equation*}

The distribution of arch-breaking times thus follows a power law with
exponent $\alpha=1+\epsilon$ for $\tilde{\tau}\rightarrow\infty$. Therefore, the unclogging transition will be reached
 by simply increasing $\epsilon$ (it takes place at $\epsilon_{c}=1$). Interestingly, the same power law tail distribution is obtained if a \emph{single} escape time $\tau$ is assigned to each energy barrier (hence, to each $k$), instead of the distribution $p(\tau|k)$ of Eq.~\ref{eq:dist_t_exp}. In this case, the description boils down to Bouchaud's trap model for aging in glasses \cite{Bouchaud1992}, in which the system hops between exponentially distributed energy traps. Related models were also devised to explain e.g. the power-law blinking of semiconductor nanocrystals \cite{verberk2002simple}.
  
We test this result against numerical simulations relying on the velocity Verlet algorithm for stochastic dynamics \cite{gronbech2013simple}. 
To this end, a particular potential well has to be specified; we have chosen 
$ V(x) = \frac{E_{b}}{2} [1 - \cos\left(\pi x\right) + e^{-\pi(1+x)} ]$ (see Fig.~\ref{fig:sketch_arch}b)
\footnote{Another potential function, $ V(x) = \frac{E_{b}}{8} [ \left(1 - \cos(\pi x)\right)^3 + 4 e^{-\pi(1+x)} ]$, was tested and was found to yield almost identical results.}.
With this specific choice, the attempt frequency $\nu$ is dependent on $E_{b}$ (via $\omega_0$ and $\omega_b$).
To account for this dependence, we invert Eq.~\ref{eq:Kramers_formula} to get
$E_b(\tau)\approx E_b^\star \epsilon \ln(\nu \tau)$, where $\nu$ is the attempt frequency for $E_b=E_b^\star$, and
we substitute $\nu\left[E_{b}\left(\tau\right)\right]$ for $\nu$ in the rescaled time $\tilde\tau = \nu \tau$.
The validity of this approach is endorsed by the coincidence between the prediction of Eq.~\ref{eq:Pcum_a=1} and the simulation results (Fig.~\ref{fig:CCDF_alphas}a).

A similar reasoning for the general case $a>0$ leads to an integral for the CCDF that cannot readily be expressed in closed form.
Still, we can resort to the approximation indicated above, i.e., neglecting fluctuations for a given trap depth $E_{b}$ and
replacing $p(\tau|k)$ with $\delta\left(\tau-k^{-1}\right)$. The approximate CCDF can then be written as
\begin{equation}
P(T>\tilde{\tau})  \approx P\left[E_{b}(T)>E_{b}(\tilde{\tau})\right]  \approx  e^{-\left(\epsilon\,\ln\,\tilde{\tau}\right)^{\nicefrac{1}{a}}}
\label{eq:CCDF_general_approximation}
\end{equation}
The numerical results obtained with the potential $V(x)$ for different values of $a$ (Fig.~\ref{fig:CCDF_alphas}b) confirm the accuracy of this expression for long time lapses and
thus support the validity of the approximation.

\begin{figure}
\includegraphics[width=\columnwidth]{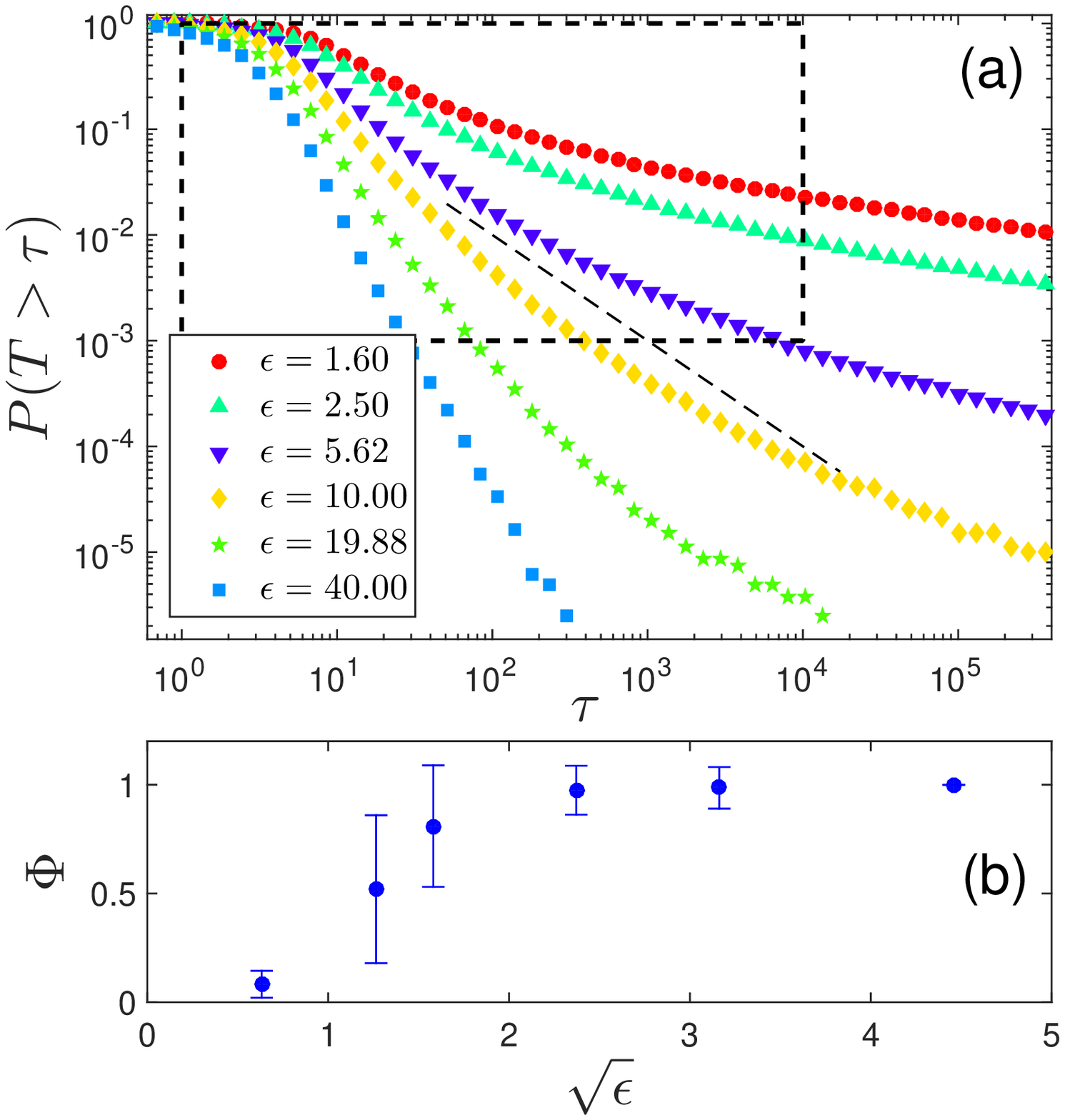}
\caption{
\label{fig:CCDF_aEq2} \textbf{(a)} Survival functions $P(T>\tau)$ as a function of dimensional time
$\tau$ for $a = 2$, $\gamma=0.1$, $E_b^\star=1$, and different $\epsilon$ as indicated in the legend. The box in thick dashed line
indicates the experimentally accessible values. The thin dashed line has slope -1. \textbf{(b)} Duty cycle $\Phi$ (see main text) calculated over a time window of $1000\,\mathrm{s}$ as a function of the vibrational acceleration $\sqrt{\epsilon} \sim \Gamma$, using the same parameters as in panel (a).
For this calculation, the flow intervals were set to $t_f=10\,\mathrm{s}$ and the model time unit was set to $1\,\mathrm{ms}$, to get closer to the experimental values of \cite{janda2009unjamming}.
The error bars represent standard deviations.}
\end{figure}

Let us now focus on the value $a=2$ which, as mentioned before, is the one that we inferred from the vibration ramp experiments. The CCDF $P(T>\tau)$ for various vibrational temperatures $\epsilon$ are plotted in Fig.~\ref{fig:CCDF_aEq2}a, as a function of non-rescaled time $\tau$. In the experimentally accessible region ($P(T>\tau)>10^{-3}$), delimited by the thick box on the figure, the survival functions are
well described by power laws with exponents that hint at a transition between a clogged regime ($\alpha\leq2$, diverging $\left\langle \tau\right\rangle$) and
an unclogged regime  ($\alpha>2$, converging $\left\langle \tau\right\rangle$), in excellent agreement with experimental findings \footnote{A more direct
comparison with experimental data is proposed in the Supplemental Material}. Experimentally, this transition was observed when increasing the vibration intensity $\Gamma$ or the outlet size. Both changes come down to increasing $\epsilon$ in the model, because $\epsilon \sim  \Gamma^{2} / \langle \Gamma_c \rangle ^2$ and the lower arch stabilities for larger outlets translate into lower typical energy barriers $E_{b}^{\star}$, hence larger $\epsilon$.
Quite interestingly, the model also captures the flattening of $P(T>\tau)$ at large $\tau$ that was observed experimentally
for weak vibrations (especially at high frequencies) or narrow apertures \cite{lozano2015stability}, i.e., at small $\epsilon$. But, remarkably,
 the model suggests that this flattening is a \emph{generic} property of gently shaken flows,
which arises because of the heavier than exponential tail in $p(E_b)$.

The flattening of the survival function has crucial implications for the unclogging transition exposed above. Indeed, we notice 
from Eq.~\ref{eq:CCDF_general_approximation} that $\left\langle \tau\right\rangle =\int_{0}^{\infty}P(T>\tau)d\tau$ diverges for any $a>1$, regardless of the vibrational temperature
$\epsilon$. Therefore, the model predicts that the system is always in the clogged regime, provided that the aperture gives rise to exponentially distributed arch stabilities $\Gamma_{c}$ (Eq.~\ref{eq:D(Eb)} with $a=2$). The reasoning holds as long as (i) there is no upper cutoff in  $p(E_b)$ in the range of experimentally relevant values and (ii) vibrations are weak enough to not affect the granular density near the outlet, thus leaving $p(E_b)$ mostly unaltered as compared to the vibrationless situation \cite{mankoc2009role}. 
On no account does this conclusion prevent experiments of \emph{finite} duration $T$ from appearing
to be in the flowing state if the vibrations are strong enough. Indeed, the duty cycle $\Phi \equiv \langle t_f \rangle / (\langle t_f \rangle + \langle \tau \rangle)$, which quantifies the fraction of time that the system spends effectively flowing, will reach finite values (intermittent flow) at high enough $\epsilon$, if it is computed  within a temporal window of finite duration
$T$. The example plotted in Fig.~\ref{fig:CCDF_aEq2}b for $T=1000\,\mathrm{s}$ is in fact very similar to the measurements by Janda \emph{et al.} (see Fig.~7 of \cite{janda2009unjamming},
where $1-\Phi$ is plotted).

Now that several features of the unclogging process in vibrated silos have been recovered, let us extend the model
to granular flows in static silos. When the system is flowing, the motion of grains creates an internal agitation, with a kinetic temperature proportional to the kinetic energy per
grain: $T_{K}\propto E_{K}$. Using this temperature in our trap model, we expect the system to escape almost immediately from shallow traps $E^{(j)}_{b}\leqslant E_{K}$, where $j=1\ldots s-1$ numbers the successive energy barriers. Only when a barrier of height $E^{(s)}_{b}>E_{K}$ is finally encountered will the system be arrested in the trap; without external agitation this halt will last forever. The clogging probability $p_c$ per grain is then the probability to encounter
such a high barrier, $p_{c} = P(E_{b}>E_{K})$.
Furthermore, under the assumption that the $E_b^{(j)}$ are uncorrelated, the number  $s$ of grains that have escaped prior to clogging follows a Bernoulli process. This naturally leads to an exponential distribution of avalanches between clogs, with a mean
size $\left\langle s\right\rangle =p_{c}^{-1}$ for $p_c\ll 1$.

In a recent numerical work, Arevalo \textit{et al.} computed the avalanche size for different values of gravity $g_{\mathrm{eff}}$ \citep{arevalo2016clogging}. Besides confirming the expected scaling $E_{K}\propto g_{\mathrm{eff}}$ and  showing that $E_K$ is a
 prominent parameter for the description of the flow, their main result is the scaling law $\ln\left\langle s\right\rangle \propto\sqrt{g_{\mathrm{eff}}}$. From these relations, we arrive at
\[
P(E_{b}>E_{K})=\exp\left(-b\sqrt{E_{K}}\right),
\]
where $b$ is a positive constant. Strikingly, the pdf derived from this CCDF takes the form
\[
p(E_{b})\sim\sqrt{\frac{E_{b}^{\star}}{E_{b}}}\exp\left(\sqrt{\frac{E_{b}}{E_{b}^{\star}}}\right). \]
This is a Weibull distribution with exactly the same shape ($a=2$) as the one we inferred
from ramp experiments in a vibrated silo. Thus, the distribution of energy barriers obtained in a vibrated silo is 
compatible with the avalanche size dependence on gravity in a static one.
%; hence suggesting that the physics behind the clogging and unclogging processes is the same.

In summary, we have put forward a model which likens unclogging at a bottleneck to the exploration of a simple energy landscape.
We derived the statistics of escape times for a generic distribution of energy barriers. For the specific distribution
inferred from measurements of arch stabilities, the escape time statistics are consistent with the heavy tails characteristic of flows through bottlenecks. An abundance of extremely long-lived clogs emerges generically in the model and it is thus 
suggested that gentle vibrations may not restore a permanent steady flow. The model is then extended to static silos, in which clogs persist forever if they can resist until the kinetic energy of all the grains is fully dissipated. We find that the variations of avalanche sizes with gravity reported recently
stem from the very same distribution of barriers as that obtained from experiments in vibrated silos. This relation challenges the widespread idea that clogging and unclogging are independent processes that require separate interpretations.

\vspace{0.7cm}

\begin{acknowledgments}
AN is funded by Centre National de la Recherche Scientifique. AG and IZ acknowledge funding from Ministerio de Econom\'{\i}a y
Competitividad (Spanish Government) through Project No. FIS2014-57325 and FIS2017-84631. AN thanks Marie Chupeau for discussions.
\end{acknowledgments}

\clearpage
\appendix
 \setcounter{table}{0}
       \renewcommand{\thetable}{S\arabic{table}}%
       \setcounter{figure}{0}
       \renewcommand{\thefigure}{S\arabic{figure}}%
       \setcounter{equation}{0}
       \renewcommand{\theequation}{S\arabic{equation}}%

\section{Determination of barrier heights from vibration ramp-up experiments}

This section exposes the detail of calculations supporting our claim
that, if stable arches (or vaults) are likened to energy traps which
detain the system, the barrier height $E_{b}$ can be determined from
ramp experiments in which the vibrational intensity $\Gamma$ is gradually
increased from zero at a rate $\dot{\Gamma}$ until the arch breaks,
which occurs at a critical value $\Gamma_{c}$.

Let $P_{s}(t)\in[0,1]$ be the survival probability of an arch up
to time $t$, with $P_{s}(t=0)=1$. At a given vibrational intensity
$\Gamma=\dot{\Gamma}t$, the arch-breaking rate is the Kramers escape
rate $\nu e^{-\beta E_{b}}$, from Eq.~2 of the main text, so the
survival probability evolves according to
\[
-\dot{P}_{s}\,dt=P_{s}\nu e^{-\beta E_{b}}dt,
\]
where $P_{s}$ is the probability that the arch has survived so far
and $\beta=\gamma/(\dot{\Gamma}t)^{2}$. Dividing both sides by $P_{s}$
and integrating up to the breaking time $t_{c}=\Gamma_{c}/\dot{\Gamma}$
yields
\begin{eqnarray*}
-\ln\left[P_{s}(t_{c})\right] & = & \nu\int_{0}^{t_{c}}\exp\left(\frac{-\gamma E_{b}}{\dot{\Gamma}^{2}t^{2}}\right)dt.
\end{eqnarray*}
Since the arch will typically break when its survival probability
is $\nicefrac{1}{2}$, after a change of integration variables from
$t$ to $u=\dot{\Gamma}t/\sqrt{\gamma E_{b}}$, one arrives at
\begin{eqnarray}
\ln(2) & = & \frac{\nu\sqrt{\gamma E_{b}}}{\dot{\Gamma}}\,F\left(\frac{\dot{\Gamma}t_{c}}{\sqrt{\gamma E_{b}}}\right),\nonumber \\
\frac{\dot{\Gamma}\ln(2)}{\nu\sqrt{\gamma E_{b}}} & = & F\left(\frac{\Gamma_{c}}{\sqrt{\gamma E_{b}}}\right)\label{eq:F_app}
\end{eqnarray}
with $F(x)=\int_{0}^{x}\exp\left(-u^{-2}\right)\,du$. At first sight,
solving Eq.~\ref{eq:F_app} for $E_{b}$ looks challenging. However,
it suffices to notice that the monotonic function $F(x)$ sharply
increases from $10^{-7}$ to $5\cdot10^{-3}$ when $x$ spans the
(narrow) range $[0.3,\,0.6]$. Experimentally, the ratio $\frac{\dot{\Gamma}\ln(2)}{\nu\sqrt{\gamma E_{b}}}$
appearing in Eq.~\ref{eq:F_app} most probably took values around
$10^{-5}$ in \cite{lozano2012breaking,lozano2014estabilidad,lozano2015stability}.
Indeed, $\sqrt{\gamma E_{b}}\sim\Gamma_{c}$ (very roughly speaking,
from the leading order of Eq.~\ref{eq:F_app}), $\dot{\Gamma}/\Gamma_{c}\approx10^{-2}$,
and it is sensible to approximate the attempt frequency $\nu$ with
the vibration frequency, of order $10^{2}-10^{3}\,\mathrm{Hz}$ (see
\emph{e}.\emph{g. }Table~II of \cite{lozano2015stability}), for
want of a more accurate value. Therefore, the argument $\left(\Gamma_{c}/\sqrt{\gamma E_{b}}\right)$
of $F$ is between 0.3 and 0.6 (and very probably between 0.35 and
0.5), so
\begin{equation}
\gamma\,E_{b}\approx6\Gamma_{c}^{2}\pm2\Gamma_{c}^{2}.\label{eq:Eb_app}
\end{equation}

We have thus shown how to relate the barrier height $E_{b}$ to the
critical vibrational intensity $\Gamma_{c}$.

\section{Comparison of the distribution of clog durations with experimental
data}

This section is aimed at providing a direct comparison between the
distributions of clog durations $\tau$ found with our trap model
and experimental ones. To this end, we consider Lozano et al.'s two-dimensional
hopper flow experiments for an aperture of width $D=4.2$ particle
diameters, in the presence of vibrations of frequency $1\,\mathrm{kHz}$.
For this frequency, critical vibration intensities $\Gamma_{c}$ (see
the previous section) were measured in ramp-up experiments for apertures
$D\in[3.2,\,4.5]$. From these measurements, shown in Fig.~3.11 (p.~72)
of \cite{lozano2014estabilidad}, we can interpolate that $\left\langle \Gamma_{c}\right\rangle \simeq1.7$
for $D=4.2$. 

\begin{figure*}
\noindent \begin{centering}
\includegraphics[width=0.49\textwidth]{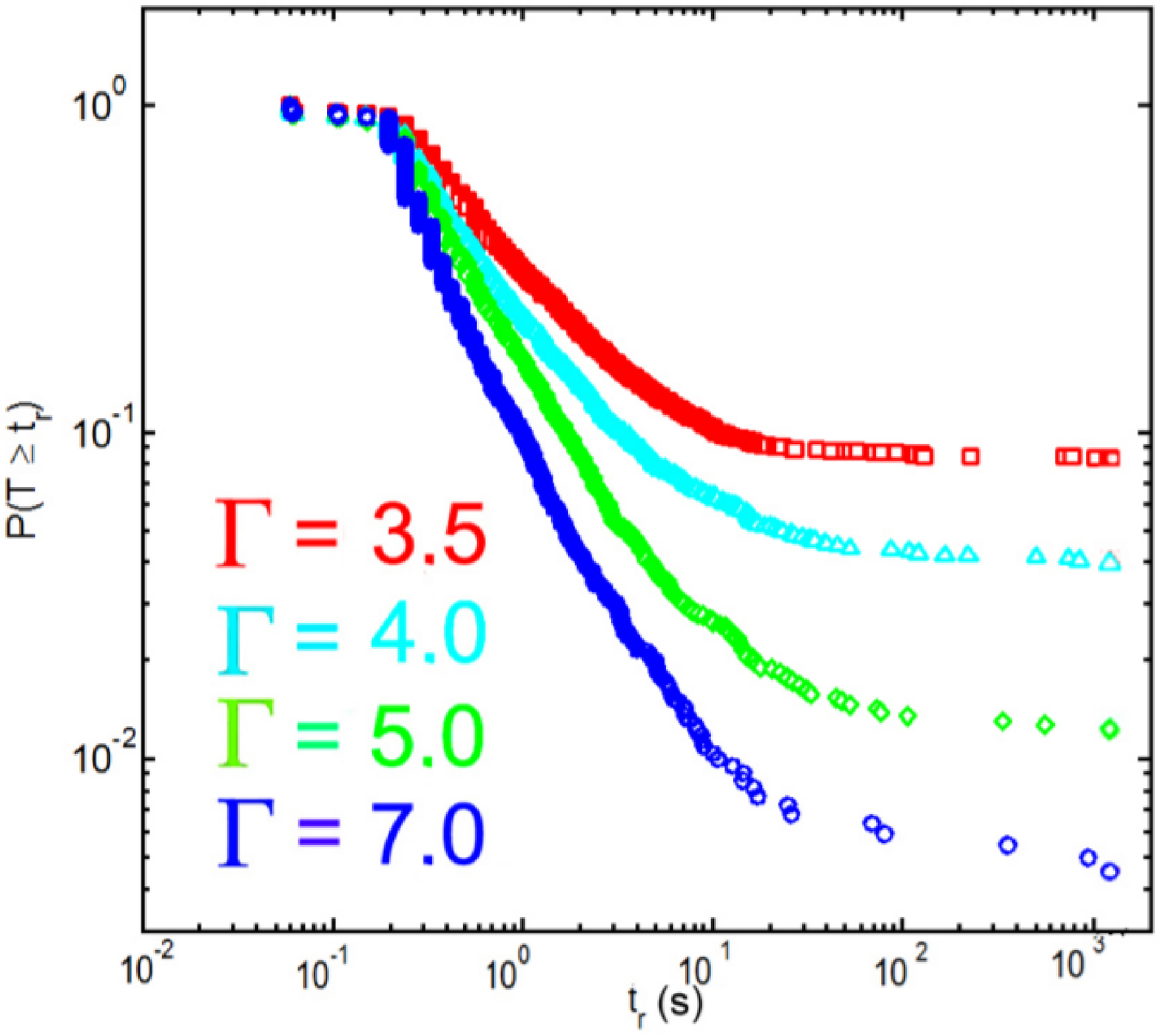}~\includegraphics[width=0.49\textwidth]{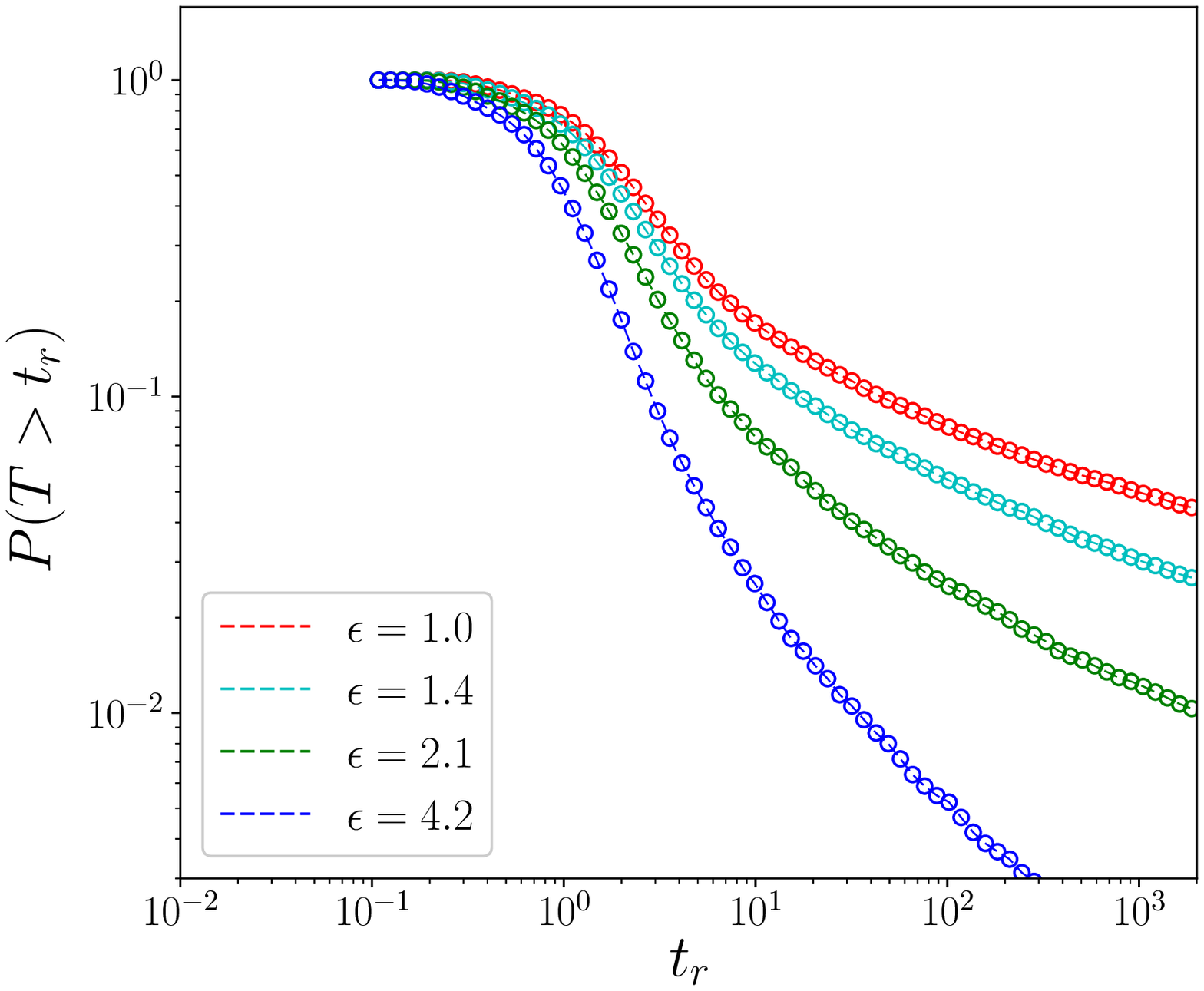}
\par\end{centering}

\caption{\label{fig:CCDF_Lozano_app}Survival function $P(T\geqslant t_{r})$
of clog durations $t_{r}$ (called $\tau$ in the main text) in Lozano's
two-dimensional experiments at a vibration frequency of $1\,\mathrm{kHz}$
and for an aperture of width $D=4.2$ (\emph{left}) and in our trap
model, with $\gamma=0.3$ and $E_{b}^{\star}=1$ (\emph{right}). The
left panel was taken from \cite{lozano2014estabilidad} (p.~101).
The model time unit was set to $0.2\,\mathrm{s}$ to facilitate the
comparison. Notice that both panels are plotted with identical axes.}
\end{figure*}

The vibrational intensities $\Gamma=3.5,\,4.0,\,5.0$, and 7.0 studied
experimentally then correspond to vibrational temperatures $\epsilon=1.0,\,1.4,\,2.1$,
and 4.2. To obtain these correspondences, we have made use of $E_{b}\approx4\Gamma_{c}^{2}/\gamma$
(where we arbitarily chose the lower bound in Eq.~\ref{eq:Eb_app},
because it gives better results) and $E_{b}=E_{b}^{\star}\Gamma_{c}^{2}/\left\langle \Gamma_{c}\right\rangle ^{2}$
(Eq.~5 of the main text), hence $\gamma E_{b}^{\star}\approx4\left\langle \Gamma_{c}\right\rangle ^{2}$
and $\epsilon\equiv\frac{\Gamma^{2}}{\gamma E_{b}^{\star}}=\frac{1}{4}\frac{\Gamma^{2}}{\left\langle \Gamma_{c}\right\rangle ^{2}}$.
The survival functions (CCDF) $P(T>\tau)$ of clog durations $\tau$
found in the model are directly compared to the experimental ones
in Fig.~\ref{fig:CCDF_Lozano_app}. Given the simplicity of our trap
model and bearing in mind that the main trends of the CCDF only depend
on one parameter, $\epsilon$, the agreement is deemed quite satisfactory,
albeit imperfect.

\begin{figure*}
\noindent \begin{centering}
\includegraphics[width=0.49\textwidth]{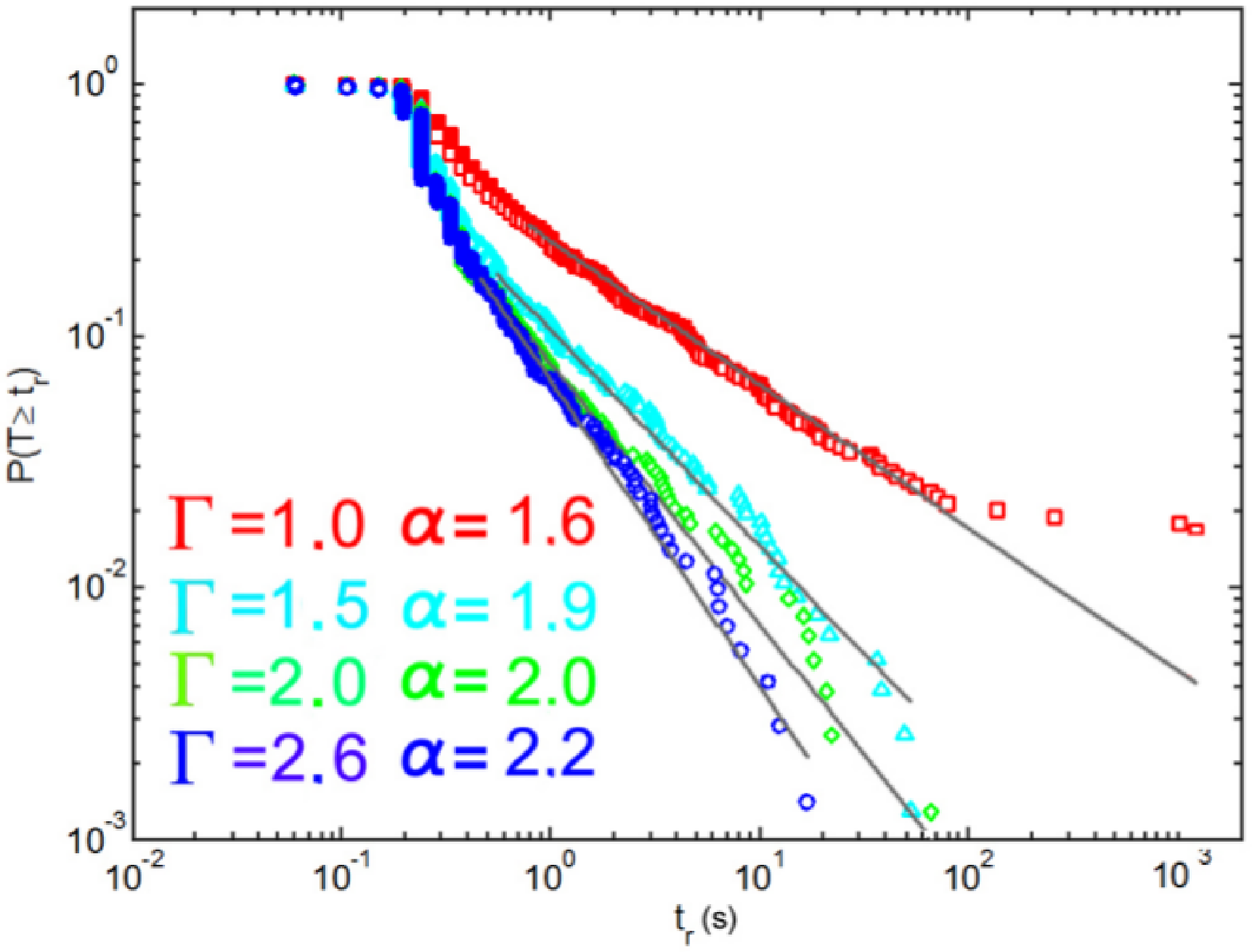}~\includegraphics[width=0.49\textwidth]{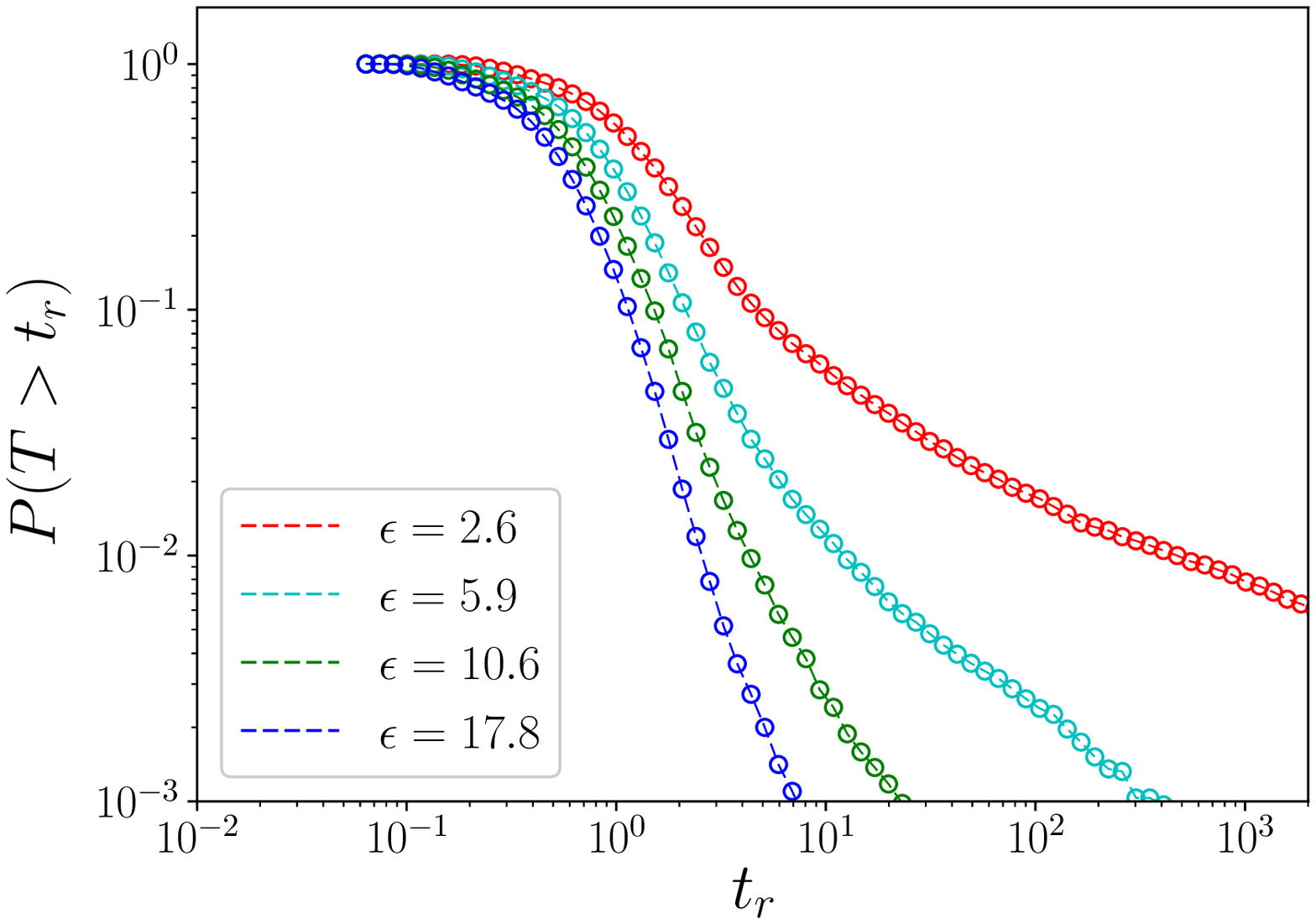}
\par\end{centering}

\caption{\label{fig:CCDF2_Lozano_app}Survival function $P(T\geqslant t_{r})$
of clog durations $t_{r}$ measured experimentally at a frequency
of $100\,\mathrm{Hz}$ and for an aperture $D=4.76$ (\emph{left})
and in our trap model, with parameters $\left(\gamma,E_{b}^{\star}\right)$
identical to Fig.~\ref{fig:CCDF_Lozano_app} (\emph{right}). The
left panel was taken from \cite{lozano2014estabilidad} (Fig.~3.28,
p.~94). The model time unit was set to $0.2\,\mathrm{s}$.}
\end{figure*}

Varying the setup, we now consider an aperture of width $D=4.76$
with vibrations of frequency 100 Hz. Unfortunately, ramp experiments
were not performed experimentally in this setup to measure $\Gamma_{c}$.
All we can do is to assume that, for this aperture, the ratio of the
mean critical values $\left\langle \Gamma_{c}\right\rangle $ for
vibration frequencies 100Hz and 1000Hz is similar to that measured
for an aperture $D=4.0$, i.e., $\frac{0.40}{1.80}\simeq0.22$, following
Table~I of \cite{lozano2015stability}. Then, we expect $\left\langle \Gamma_{c}\right\rangle =0.31$
at a frequency of 100Hz for $D=4.76$. It follows that vibrational
intensities $\Gamma=1.0,$ 1.5, 2.0, and 2.6 correspond to $\epsilon=2.6,$
5.9, 10.5, and 17.8. Figure~\ref{fig:CCDF2_Lozano_app} presents
the comparison between the experimental CCDF and the modeled CCDF.
Once again, we find satisfactory agreement.

\end{document}